\documentclass[11pt]{article}
\usepackage{amssymb}
\usepackage{color}
\begin{document}
\title{\textbf{\textbf{Noncommutative Phase spaces on Aristotle group}}}
\author{Ancille Ngendakumana\footnote{nancille@yahoo.fr}\\Institut de
Math\'ematiques et des Sciences Physiques, Porto-Novo, B\'enin\\ and \\
Joachim Nzotungicimpaye\footnote{kimpaye@kie.ac.rw}
\\Kigali Institute of Education, Kigali, Rwanda\\ and \\ Leonard Todjihound\'e\footnote{leonardt@imsp.uac.org}\\Institut de
Math\'ematiques et des Sciences Physiques, Porto-Novo, B\'enin\\}

\maketitle
\date
\begin{abstract}
 We realize noncommutative phase spaces as coadjoint orbits of \\extensions of the Aristotle group
in a two-dimensional space.  Through these constructions the momenta of the phase spaces
do not commute due to the presence of a naturally introduced magnetic field.
These cases correspond to the minimal coupling of the momentum with a magnetic potential.
\end{abstract}
\section{Introduction}
Classical electromagnetic interaction can be introduced through the modified symplectic form $\sigma=dp_i\wedge
dq^i+\frac{1}{2}F_{ij}dq^i\wedge dq^j$ (\cite{abraham}, \cite{guillemin}, \cite{souriau}).  This has been
initiated by J.M.Souriau (\cite{souriau}) in the seventies : one of his important theorems says in fact that when a symmetry group $G$
acts on a phase space transitively, then the latter is a coadjoint orbit of $G$, endowed with its canonical symplectic form.
 The original applications that Souriau presented in his book (\cite{souriau}) concern
both the Poincar\'e and the Galilei groups.
 Recently many authors (\cite{horvathy12}, \cite {horvathy11},  \cite{horvathy},
\cite{horvathy01}, \cite{vanhecke11},...) generalized
the modification of the symplectic form by introducing the so-called dual magnetic field $G$ by considering $\sigma=dp_i\wedge
dq^i+\frac{1}{2}F_{ij}dq^i\wedge dq^j+\frac{1}{2}G^{ij}dp_i\wedge dp_j$.  The fields $F$ and $G$ are responsible of the
noncommutativity respectively of momenta and positions.
Noncommutative phase spaces are then defined as spaces on which coordinates satisfy the relations:
 \begin{eqnarray*}\label{poisson1}
 \{ q^i,q^j \}=G^{ij}~~,~~   \{ q^i,p_j \}=\delta^{i}_j~~,~~ \{ p_i,p_j \}= F_{ij}
\end{eqnarray*}
where $\delta^{i}_j$ is a unit matrix, whereas $G^{ij}$ and $F_{ij}$ are functions of positions and momenta.
Moreover the physical dimensions of $G^{ij}$ and $F_{ij}$ are respectively $M^{-1}T$ and $MT^{-1}$, $M$ representing a mass while $T$
represents a time. \\

In more recent times, Souriau's ideas were later extended to other groups.  In (\cite{duval1}) for example, a classical \textquotedblleft photon
\textquotedblright model was constructed, based entirely on the Euclidian group $E(3)$.  As the latter is simultaneously a subgroup of both the
Poincar\'e and the Galilei groups, hence the \textquotedblleft euclidian photon\textquotedblright constructed by Souriau's  orbit method is indeed
a reduction of both the relativistic and the nonrelativistic massless models as presented by Souriau (\cite{souriau}).
There is an intermediate group between the Euclidian and the Galilei groups dubbed, again by Souriau (\cite{souriau1}), the Aristotle group :
it also contains time translations but not boosts. \\ This work is precisely to study the classical dynamical systems associated with this
intermediate group.
We use Souriau's method also called coadjoint orbit method to contruct phase spaces endowed with modified
symplectic structure on the Aristotle group.  Explicitly, we demontrate that such deformed objects can be generated in the framework of
noncentrally extended Aristotle algebra
as well as in the framework of its corresponding central extension.  The obtained in such a way phase spaces do not commute in momentum sector
due to the presence of a naturally introduced magnetic field. In other words, the obtained cases correspond to the minimal coupling of the momentum
with a magnetic potential.

Note that there has been other more recent works about a similar construction starting with the centrally extended \textquotedblleft
anisotropic Newton-Hooke\textquotedblright \\groups (\cite{ancilla}) and with the noncentrally extended of both Para-Galilei and
 Galilei groups (\cite{ancilla1}) in a two-dimensional space.\\

The paper is organized as follows. In section two, we give symplectic realizations of the Aristotle group in two-dimensional space using its
first and second central extensions. In the third section, we realize symplectically both the noncentrally extended
Aristotle Lie group and its central extension counterpart.  As the coadjoint orbit construction
has not been curried through this Lie group before, physical interpretations of new generators of the extended
corresponding Lie algebras are also given.

 \section{First and second central extensions of the \\Aristotle group}
It is well known that a free dynamical system is a geometric object for the Aristotle group (\cite{souriau1}) which is
the group of both Euclidean displacements and time translations.  Explicitly, the Aristotle group $A(2)$ in a two-dimensional
space is a Lie group whose multiplication law is given by
\begin{eqnarray}\label{Aristotlelaw}
(\theta, \vec{x}, t)(\theta^{\prime}, \vec{x}^{~\prime}, t^{\prime})=(\theta+\theta^{\prime}, R(\theta)\vec{x}^{~\prime}+\vec{x}
 , t+t^{\prime})
\end{eqnarray}
where $\vec{x}$ is a space translation vector , $t$ is a time
translation parameter and $\theta$ is a rotation parameter.  \\
Its Lie algebra $\cal{A}$ is then generated by
the left invariant vector fields
\begin{eqnarray*}
J=\frac{\partial}{\partial\theta},~~\vec{P}=R(-\theta )\frac{\partial}{\vec{x}},~~H=\frac{\partial}{\partial t}
\end{eqnarray*}
such that the only nontrivial Lie brackets are
\begin{eqnarray}\label{Aristotleliealgebra}
[J,P_i]=P_j\epsilon^j_i,~~i,j=1,2.
\end{eqnarray}
The multiplication law (\ref{Aristotlelaw}) implies that the element $g$ of this group can be written as:
\begin{eqnarray}\label{elementg}
g=\exp(\vec{x}\vec{P}+t H)\exp(\theta J)
\end{eqnarray}

\subsection{First central extension of $A(2)$}
From the relation  $\exp(2\pi J)H \exp(-2\pi J)=H$ and by use of the standard methods (\cite{hamermesh}, \cite{kirillov}, \cite{kostant},
\cite{5bacry},
 \cite{nzo1}), we obtain the following nontrivial Lie brackets for the first central extension ${\hat{\cal{A}}}$ of ${\cal{A}}(2)$
\begin{eqnarray}\label{centralliebrackets}
[J,P_i]=P_j\epsilon^j_i,~~ [P_i,P_j]=\frac{1}{r^2}S\epsilon_{ij},~~i,j=1,2
\end{eqnarray}
where $S$ generates the center of  ${\hat{\cal{A}}}$ while $r$ is a constant whose dimension is a length. \\
Let $g$ be given by (\ref{elementg}) and ${\hat g}=\exp(\varphi S)g$ be the corresponding element in the connected Lie group associated to
the extended Lie algebra ${\hat{\cal{A}}}$.   By use of the Baker-Campbell-Hausdorff formulae (\cite{bchformulae}) and by identifying ${\hat g}$ with
$(\varphi,\theta,\vec{x},t)$, we find that the multiplication law of the connected extended Lie group is:
\begin{eqnarray*}
(\varphi,\theta,\vec{x},t)(\varphi^{\prime},\theta^{\prime},\vec{x}^{~\prime},t^{\prime})=(\varphi^{\prime}+
\frac{1}{2r^2}R(-\theta)\vec{x}\times \vec{x}^{~\prime}+\varphi,
\theta+\theta^{\prime}, R(\theta)\vec{x}^{~\prime}+\vec{x},t+t^{\prime})
\end{eqnarray*}
or equivalently
\begin{eqnarray*}
 (\alpha,g)(\alpha^{\prime},g^{\prime})=(\alpha+\alpha^{\prime}+c(g,g^{\prime}),g g^{\prime})
\end{eqnarray*}
where $c(g,g^{\prime})=\frac{1}{2r^2}R(-\theta)\vec{x}\times \vec{x}^{~\prime}$
is a two-cocycle and $ g g^{\prime}$ is the multiplication law (\ref{Aristotlelaw}).\\
The adjoint action $Ad_g(\delta{\hat{g}})=g( \delta{\hat{g}} )g^{-1}$ of $A$ on the Lie algebra ${\hat{\cal{A}}}$ is given by:
\begin{eqnarray*}\label{ancillaa}
 Ad_{(\theta,\vec{x},t)}(\delta \varphi,\delta \theta,\delta \vec{x},\delta t)=(\delta \varphi+\frac{1}{r^2}R(-\theta)\vec{x}\times \delta \vec{x}-
\frac{1}{2r^2} \vec{x}~^2\delta\theta,\delta \theta,R(\theta)\delta \vec{x} +\epsilon(\vec{x}) \delta\theta, \delta t)
\end{eqnarray*}
with
\begin{eqnarray}\label{epsilon}
\epsilon(\vec{x})=\left(
\begin{array}{cc}
0&x^{2}\\-x^{1}&0
\end{array}
\right)
\end{eqnarray}
If the duality between the extended Lie algebra and its dual is given by the action
$j\delta \theta+\vec{p}.\delta \vec{x}+l\delta \varphi+E\delta t$,
where $\vec{p}$ is a linear momentum, $l$ is an action, $j$ is an angular momentum while $E$
is an energy, then the coadjoint action of the Aristotle Lie group is
\begin{eqnarray}\label{coadjointaction1}\nonumber
Ad^*_{(\vec{x},t,\theta)}(j,\vec{p},l,E)=(j+\frac{m\omega}{2}(\vec{x}~^2)+\vec{x} \times R(\theta)\vec{p}, R(\theta)\vec{p}-m\omega\epsilon(
{\vec{x}})
,l,E)
\end{eqnarray}
where we have used the "wave-particule duality" $l\omega=mc^2$ and the relation $c=\omega r$ linking the velocity $c$, the frequency $\omega$
and the universe radius $r$.\\
The Kirillov form in the basis
$(J,P_1,P_2,H,S)$ is
\begin{eqnarray*}
K(a)=\left(
\begin{array}{ccccc}
0&p_2&-p_1&0&0\\
-p_2&0&m\omega&0&0\\
p_1&-m\omega&0&0&0\\
0&0&0&0&0\\
0&0&0&0&0\\
\end{array}
 \right)
\end{eqnarray*}

The coadjoint orbit of the central extended Lie group on the dual of its Lie algebra is characterized by the two trivial invariants $l$ and $E$,
and a nontrivial invariant
\begin{eqnarray}\label{Aristotleinvariant10}
s=j+\frac{p^2}{2m\omega}+\frac{m\omega{q}^2}{2}
\end{eqnarray}
where $p$ and $q$ are defined by
\begin{eqnarray}\label{ancilla0}
 q=-\frac{p_2}{m \omega},  ~~~p=p_1
\end{eqnarray}
 Let us denote by  ${\cal{O}}_{(s,l,E)}$ the maximal coadjoint orbit of the Aristotle group $A(2)$ on the dual of its central extended Lie
algebra.\\
The restriction $\Omega=(\Omega_{ab})$ of the Kirillov form to the orbit is then
\begin{eqnarray*}
\Omega=\left(
\begin{array}{cc}
0&m\omega \\-m\omega&0\\
\end{array}
\right)
\end{eqnarray*}
 It follows that the symplectic form is then $\sigma=dp\wedge dq$.\\
The symplectic realization of the Aristotle Lie group is
\begin{eqnarray*}
 D_{(\theta,\vec{x},t)}(p,q)=(\cos \theta ~p+m\omega q\sin\theta-m\omega x^2,-\frac{p}{m\omega}\sin\theta+q\cos\theta-x^1)
\end{eqnarray*}
The Poisson bracket  corresponding to this symplectic structure is then
the canonical one and the time translation subgroup acts trivially on the orbit.\\
To overcome this fact, let us study the symplectic realization of the second central extended Aristotle Lie group.
\subsection{ Second central extension }
 By using standard methods, we have that the second central extension of Aristotle Lie algebra in two-dimensional space
is generated by: $J,P_1,P_2,H,S,N$ satisfying the nontrivial Lie brackets:
\begin{eqnarray*}
[J,P_i]=P_j\epsilon^j_i, ~~[ P_i,P_j]=\frac{1}{r^2}S\epsilon_{ij},~~~[S,H]=\omega N
\end{eqnarray*}
The multiplication law of this extended Lie group is given by:\\
$(\psi,\varphi,\theta,\vec{x},t)(\psi^{\prime},\varphi^{\prime},\theta^{\prime},\vec{x}^{~\prime},t^{\prime})$
\begin{eqnarray*}
=(\psi+\psi^{\prime}-\omega t\varphi^{\prime},\varphi^{\prime}+
\frac{1}{2r^2}R(-\theta)\vec{x}\times \vec{x}^{~\prime}+\varphi,
\theta+\theta^{\prime}, R(\theta)\vec{x}^{~\prime}+\vec{x},t+t^{\prime})
\end{eqnarray*}
 The adjoint action of ${\hat{A}}$ on its extended Lie algebra is explicitly given by \\
$Ad_{(\varphi,\theta,\vec{x},t)}(\delta \psi,\delta\varphi,\delta \theta,\delta\vec{x},\delta t)$
\begin{eqnarray*}
=(\delta \psi-\omega t\delta \varphi+\omega \varphi
\delta t,\delta \varphi+\frac{1}{r^2}R(-\theta)\vec{x}\times \delta \vec{x}-
\frac{1}{2r^2} \vec{x}^2\delta\theta,\delta\theta,R(\theta)\delta \vec{x}+\epsilon(\vec{x})\delta \theta,\delta t)
\end{eqnarray*}
where $\epsilon (\vec{x})$ is given by the relation (\ref{epsilon}).\\
If the duality between the extended Lie algebra and its dual is given by the action
$j\delta \theta+\vec{p}.\delta \vec{x}+E\delta t+l\delta \varphi+h\delta\psi$,
 then the coadjoint action of the extended Aristotle Lie group is such that \\
$ Ad^*_{(\varphi,\vec{x},t,\theta)}(j,\vec{p},E,l,h)$
\begin{eqnarray}\label{coadjointaction12}
=(j+\vec{x} \times R(\theta)\vec{p}-\frac{l+h\omega t}{2 r^2}{\vec{x}}~^2,
R(\theta)\vec{p}+\frac{l}{r^2}\epsilon({\vec{x}})+\frac{h}{r^2} \omega t\epsilon({\vec{x}}),E-h \omega \varphi,l+h \omega t,h)
\end{eqnarray}
meaning that $h$ is a trivial invariant.\\
In the basis
$(J,P_1,P_2,H,S,N)$, the Kirillov form is
\begin{eqnarray}\label{kirillov0}
K(a)=\left(
\begin{array}{cccccc}
0&p_2&-p_1&0&0&0\\
-p_2&0&m\omega &0&0&0\\
p_1&-m\omega &0&0&0&0\\
0&0&0&0&-h\omega&0\\
0&0&0&h\omega&0&0\\
0&0&0&0&0&0
\end{array}
 \right)
\end{eqnarray}
The coadjoint orbit of ${\hat{A}}$ on the dual ${\hat{\cal{A}}^*}$ of the second central extension Lie algebra is
characterized by the trivial invariant $h$ and by the
nontrivial invariant given by relation (\ref{Aristotleinvariant10}).\\
The maximal coadjoint orbit is quadri-dimensional and is denoted by ${\cal{O}}_{(h,s)}$.\\
The restriction $\Omega=(\Omega_{ab})$ of the Kirillov form (\ref{kirillov0}) to the orbit is then
\begin{eqnarray*}
\Omega=\left(
\begin{array}{cccc}
0&m\omega&0&0\\-m\omega&0&0&0\\0&0&0&-h \omega\\0&0&h\omega&0
\end{array}
\right)
\end{eqnarray*}
It follows that the symplectic form is in this case given by
\begin{eqnarray*}
\sigma=dp\wedge dq+d\alpha\wedge dl
\end{eqnarray*}
where
\begin{eqnarray*}\label{beta}
\alpha=\frac{E}{h\omega}
\end{eqnarray*}
\\
 From the relations (\ref{coadjointaction12}), we get that the symplectic realization of the extended group on its maximal coadjoint orbit
$(p^{~\prime},q^{~\prime},l^{\prime},\alpha^{\prime})=D_{(\varphi,\theta,x_1,x_2,t)}(p,q,l,\alpha)$ is such that
\begin{eqnarray*}
p^{\prime}=\cos\theta~p-m\omega\sin\theta~q-m\omega x^2+\frac{h}{r^2}\omega x^2t,~~l^{\prime}=l+h\omega t\\
q^{~\prime}=\frac{1}{m\omega}\sin\theta ~p+\cos \theta ~q -x^1-\frac{h}{m r^2}x^1t,~~\alpha^{\prime}=\alpha +\varphi~~~~\\
\end{eqnarray*}
The Poisson bracket of two functions $f_1$ and $f_2$ on the orbit corresponding to the above symplectic form
 is
\begin{eqnarray*}
\{f_1,f_2\}= \frac{\partial f_1}{\partial p}\frac{\partial f_2}{\partial
q}- \frac{\partial f_1}{\partial q}\frac{\partial f_2}{\partial
p}+\frac{\partial f_1}{\partial l}\frac{\partial f_2}{\partial
\alpha}-\frac{\partial f_1}{\partial \alpha}\frac{\partial f_2}{\partial
l}
\end{eqnarray*}
It follows that
$$\{p,q\}=1~~,~~~\{l,\alpha\}=1$$
the other Poisson brackets being trivial.\\ The equations of motion are then
\begin{eqnarray*}
 \frac{dp}{dt}=0~,~\frac{dl}{dt}=h\omega~,\frac{dq}{dt}=0~,~\frac{d\alpha}{dt}=0
\end{eqnarray*}
In this case, the coadjoint orbit is a direct product of two $2$-dimensional phase spaces  $R^2=\{(p,q)\})$ and  $R^2=\{(l,\alpha)\})$. Note
that $\alpha$ is a dimensionless quantity. For its particular value $\alpha=\frac{1}{2\pi}$, the energy $E$ is given by
\begin{eqnarray*}\label{energy}
 E=\hbar \omega
\end{eqnarray*}
 where $h=2\pi\hbar $, relation analogue to that of Quantum  Mechanics.\\
With the second central extension of Aristotle group in two-dimensional space, we have then also
realized a phase space with commuting coordinates (canonical case) .  Moreover, the position and the linear momentum do not depend on time.\\
We prove, in the following section, that noncommutative phase spaces can be obtained by considering the noncentral extension of the two-dimensional
Aristotle group.
\section{Noncentral extension of Aristotle group}
In the previous section, we find that one can not construct noncommutative phase spaces by coadjoint orbit method on the first and second
central extensions of the Aristotle group because symplectic structures obtained are canonical which means that positions commute as well
as momenta.\\In this section, we see that this construction is possible when we consider a noncentral extension of
this Lie group.
\subsection{Noncentrally extended group and its maximal \\coadjoint orbit}
 Let ${\hat{\cal{A}}}_1$ be the noncentrally extended Aristotle Lie algebra satisfying the non trivial Lie brackets
\begin{eqnarray}\label{noncentrallyliealgebra}
[J,P_i]=P_j\epsilon^j_i,~
[J,F_i]=F_k\epsilon^k_i,~[P_i,P_j]=\frac{1}{r^2}S\epsilon_{ij},~[P_i,H]=F_i,~i,j=1,2.
\end{eqnarray}
 If $\hat{g}=\exp(\varphi S+\vec{\eta}\vec{F})\exp(\vec{x}\vec{P})\exp(\theta J)\exp(tH)$
is the general element of the connected extended Aristotle group, we verify that the corresponding multiplication law is
\\ $
(\varphi^{\prime\prime},\theta^{\prime\prime},\vec{\eta}^{~\prime\prime},
\vec{x}^{~\prime\prime},t^{\prime\prime})=(\varphi,\theta,\vec{\eta},\vec{x},t)(\varphi^{\prime},
\theta^{\prime},\vec{\eta}^{~\prime},\vec{x}^{~\prime},t^{\prime})
$ \\with
\begin{eqnarray*}
\varphi^{\prime\prime}=\varphi^{\prime}+\frac{1}{2r^2}R(-\theta)\vec{x}\times \vec{x}^{~\prime}+\varphi,~
\vec{\eta}^{~\prime\prime}=R(\theta)\vec{\eta}^{~\prime}-R(\theta)\vec{x}^{~\prime}t+\vec{\eta}\\
\vec{x}^{~\prime\prime}=R(\theta)\vec{x}^{~\prime}+\vec{x}~,~
\theta^{\prime\prime}=\theta^{\prime}+\theta~,~t^{\prime\prime}=t^{\prime}+t
\end{eqnarray*}
It follows that the adjoint action of the noncentral extended Aristotle group on its Lie algebra is such that
\begin{eqnarray*}
\delta \theta^{\prime}=\delta{\theta},~~\delta t^{\prime}=\delta t,~~\delta \vec{x}^{~\prime}=R(\theta)\delta \vec{x} +
\epsilon(\vec{x})\delta \theta\\
~\delta\vec{\eta}^{~\prime}=R(\theta)\delta \vec{\eta}+\epsilon(\vec{\eta})\delta\theta- R(\theta)\delta \vec{x}~t
+\vec{x}\delta t~~~~~~\\
\delta \varphi^{\prime}=\delta \varphi+\frac{1}{r^2} R(\theta)\vec{x}\times \delta \vec{x}-\frac{\vec{x}^2}{2r^2}\delta \theta~~~
\end{eqnarray*}
If the duality between the extended Lie algebra and its dual is given by the action
$j\delta \theta+\vec{f}.\delta \vec{\eta}+\vec{p}.\delta \vec{x}+h\delta \varphi+E\delta t$, then the coadjoint
action is such that $h^{\prime}=h$ and
\begin{eqnarray}\label{Aristotlemomentum}
\vec{f}^{~\prime}=R(\theta)\vec{f},~\vec{p}^{~\prime}=R(\theta)\vec{p}+R(\theta)\vec{f}~t+\frac{h}{r^2}\epsilon(\vec{x})
\end{eqnarray}
\begin{eqnarray}\label{Aristotleangularmomentum}
j^{~\prime}=j+\vec{x}\times R(\theta)\vec{p}+\vec{\eta}\times R(\theta)\vec{f}-\frac{h}{r^2}{\vec{x}~^2}
\end{eqnarray}
\begin{eqnarray*}\label{Aristotleenergy}
E^{~\prime}=E-\vec{x}.R(\theta)\vec{f}
\end{eqnarray*}
The coadjoint orbits denoted by ${\cal{O}}_{(h,f,U)}$ are characterized by the trivial invariant $h$ and
 by two nontrivial invariants $f$ and $U$ given by:
\begin{eqnarray*}
f=||\vec{f}||,~~U=E+\frac{1}{m\omega}(\vec{p}\times\vec{f})
\end{eqnarray*}
where the wave-particle duality and the relation $c=\omega r$ have been used.\\
Let $f_1=f cos\phi~,~f_2=f sin\phi$.
The inverse of the restriction of the Kirillov form on the coadjoint orbit in the basis $(J, F_1, P_1, P_2)$ is
\begin{eqnarray*}
\Omega^{-1}=\frac{1}{m\omega f \sin\phi}\left (\begin{array}{cccc}
0&-m\omega&0&0\\m\omega&0&p_1&p_2 \\0&-p_1&0&-f\sin\phi\\0&-p_2&f\sin\phi&0\\
\end{array}
\right)
\end{eqnarray*}
The symplectic form on the orbit is then
\begin{eqnarray}\label{symplecticformnc1}
\sigma=dj\wedge d\phi+dp\wedge dq+\frac{p}{m\omega}dp\wedge d\phi+m\omega q~dq\wedge d\phi
\end{eqnarray}
where $p_1$ and $q$ are given by (\ref{ancilla0}).\\ The invariant $U$ can be written as $U=E+v(p\sin\phi+m\omega~q\cos\phi)$ where $v=\frac{f}{m\omega}$
is a velocity.\\
The Poisson bracket of two functions $g_1$ and $g_2$ corresponding to the symplectic form (\ref{symplecticformnc1}) is given by
\begin{eqnarray*}
\{g_1,g_2\}=\frac{\partial g_1}{\partial p}\frac{\partial g_2}{\partial q}-\frac{\partial g_1}{\partial q}\frac{\partial g_2}{\partial p}+
\frac{\partial g_1}{\partial j}\frac{\partial g_2}{\partial \phi}-\frac{\partial g_1}{\partial \phi}\frac{\partial g_2}{\partial j}\\
+m\omega q (\frac{\partial g_1}{\partial j}\frac{\partial g_2}{\partial p}-\frac{\partial g_1}{\partial p}\frac{\partial g_2}{\partial j})
-\frac{p}{m\omega}(\frac{\partial g_1}{\partial j}\frac{\partial g_2}{\partial q}-\frac{\partial g_1}{\partial q}\frac{\partial g_2}{\partial j})
\end{eqnarray*}
We then have the following non trivial Poisson brackets within the coordinates on the maximal coadjoint orbit:
\begin{eqnarray}\label{pb1}
\{j,p\}=m\omega~q~,~\{\phi,q\}=0
\end{eqnarray}
\begin{eqnarray}\label{pb2}
\{j,\phi\}=1~,~\{p,q\}=1
\end{eqnarray}
\begin{eqnarray}\label{pb3}
\{j,q\}=-\frac{p}{m\omega}~,\{\phi,p\}=0
\end{eqnarray}
The relations (\ref{pb1}) mean that momenta $(j,p)$ do not commute but the configurations coordinates $(\phi,q)$ commute, the
relations (\ref{pb2}) mean that $j$ is conjugated to $\phi$ and that $p$ is conjugated to $q$, the relations (\ref{pb3}) mean that $j$
do not commute with $q$ while $p$ commute with $\phi$. \\

 Let the symplectic realization of the extended Aristotle Lie group on its coadjoint orbit be given by
$(j^{~\prime},\phi^{~\prime}, p^{~\prime},q^{\prime})=L_{(\theta,\eta^1,\eta ^2,x^1,x^2,t})(j,\phi,p,q)$ .
By using relations (\ref{Aristotlemomentum}) and (\ref{Aristotleangularmomentum}), we obtain
\begin{eqnarray*}
j^{~\prime}=j+p(\sin\theta~x^1-\cos\theta ~x^2)-m\omega~q(\cos\theta~x^1+\sin\theta~x^2)+
\vec{\eta}\times R(\theta)\vec{f}-m\omega\vec{x}^2\\~p^{\prime}=\cos\theta~p+m\omega\sin\theta~q-m\omega x^2,~
q^{\prime}=-\frac{1}{m\omega}\sin\theta~p+\cos\theta~q+x^1,~\phi^{\prime}=\phi+\theta
\end{eqnarray*}
It follows that $L_{(0,0,0,0,0,t})(j,\phi,p,q)=(j,\phi,p,q)$ meaning that all the coordinates  $j, \phi ,p$ and $q$ on the maximal
coadjoint orbit are constant with respect to the time $t$.  To overcome this situation, let us consider the central extension of the
noncentrally extended Aristotle group.\\
But first note also that the symplectic form (\ref{symplecticformnc1}) can be written in the canonical way as
\begin{eqnarray*}\label{symplecticformnc2}
\sigma=dH\wedge d\tau+dp\wedge dq
\end{eqnarray*}
where $H=j\omega+\frac{p^2}{2m}+\frac{m\omega^2q^2}{2}$ is an energy while $\tau=\frac{\alpha}{\omega}$ is a time.\\
\subsection{Central extension of the noncentrally extended \\ Aristotle group }
Consider the central extension of the Lie algebra defined by (\ref{noncentrallyliealgebra}) satisfying the non trivial Lie brackets
 \begin{eqnarray}\label{Aristotleextc}
[J,P_j]=P_i\epsilon^{i}_j,~~~[P_i,P_j]=\frac{1}{r^2}S\epsilon_{ij}~~~~~~~~~~~~~~~\\\nonumber
[J,F_j]=F_i\epsilon^{i}_j~,~[P_i,H]= F_i,~~
[P_i,F_j]= K\delta_{ij}
 \end{eqnarray}
We recover the Lie algebra defined by (\ref{Aristotleliealgebra}) when $F_i=0$,~$K=0$ and $S=0$, the Lie algebra defined by
(\ref{centralliebrackets}) when
$F_i=0$,~$K=0$ and the Lie algebra defined by (\ref{noncentrallyliealgebra}) when $K=0$.
Consider now the general Lie algebra defined by (\ref{Aristotleextc}).\\
 Let $\hat{g}=\exp(\varphi S+\gamma K)\exp(t H)\exp(\vec{\eta}\vec{F}+\vec{x}\vec{P})\exp(\theta J)$
be the general element of the corresponding connected extended Aristotle group.  By identifying  $\hat{g}$ with $(\varphi,\gamma,t,
\vec{\eta},\vec{x},\theta)$
the multiplication law $\hat{g}^{\prime\prime}=\hat{g}\hat{g}^{\prime}$ is such that
\begin{eqnarray*}
\varphi^{\prime\prime}=\varphi^{\prime}+\frac{1}{2r^2}\vec{x}\times R(-\theta) \vec{x}^{~\prime}+\varphi,~~~\theta^{\prime\prime}=
\theta^{\prime}+\theta,~t^{\prime\prime}=t+t^{\prime},~~~~~~~~~~~~~\\
\gamma^{\prime\prime}=\gamma^{~\prime}+\frac{ 1}{2}\vec{x}. R(\theta)\vec{\eta}^{~\prime}-\frac{1}{2}(\vec{\eta}+\vec{x}~t^{\prime})
.R(\theta)\vec{x}^{~\prime}+\gamma,~~~~~~~~~~~~~~~~~~~~~~~~\\
\vec{\eta}^{~\prime\prime}=R(\theta)\vec{\eta}^{~\prime}+\vec{\eta}+\vec{x}~t^{\prime},~~\vec{x}^{~\prime\prime}=
R(\theta)\vec{x}^{~\prime}+\vec{x}~~~~~~~~~~~~~~~~~~~~~~~~~~~~~~~~\\
\end{eqnarray*}
It follows that the adjoint action of the extended Aristotle group on its Lie algebra is such that
\begin{eqnarray*}
\delta \gamma^{\prime}=\delta \gamma+\vec{x} \times R(\theta)\delta \vec{\eta}-\vec{\eta}\times R(\theta)\delta \vec{x}-
\vec{\eta}\times\vec{x}\delta{\theta}
+\frac{1}{2}\vec{x}^2\delta t~\\
\delta\vec{\eta}^{~\prime}=R(\theta)\delta \vec{\eta}+\epsilon(\vec{\eta}-\vec{x}~t)\delta\theta-t R(\theta)\delta \vec{x}+\vec{x}\delta t\\
\delta \varphi^{\prime}=\delta \varphi+\frac{1}{r^2} R(-\theta)\vec{x}\times \delta \vec{x}-\frac{\vec{x}^2}{2r^2}\delta \theta\\
\delta \vec{x}^{~\prime}=R(\theta)\delta \vec{x} +\epsilon(\vec{x})\delta \theta,~\delta t^{\prime}=\delta t,~\delta \theta^{\prime}=\delta{\theta}
\end{eqnarray*}
where $\epsilon (\vec{x})$ is given by the relation (\ref{epsilon}).\\
If the duality between the extended Lie algebra and its dual Lie algebra gives rise to the action
$j\delta \theta+\vec{f}.\delta \vec{\eta}+\vec{p}.\delta \vec{x}+h\delta \varphi+E\delta t+k\delta\gamma$, then the coadjoint action is such that
\begin{eqnarray}\label{massaction0}
h^{\prime}=h,~~k^{\prime}=k
\end{eqnarray}
and
\begin{eqnarray}\label{Aristotlemomentum0}
\vec{p}^{~\prime}=R(\theta)\vec{p}+R(\theta)\vec{f}~t+k(\vec{\eta}-\vec{x}~t)+\frac{h}{r^2}\epsilon(\vec{x})~,\vec{f}^{~\prime}=
R(\theta)\vec{f}-k\vec{x}
\end{eqnarray}
\begin{eqnarray*}\label{Aristotleangularmomentum0}
j^{~\prime}=j+\vec{x}\times R(\theta)\vec{p}+\vec{\eta}\times R(\theta)\vec{f}-\frac{h}{2r^2}{\vec{x}~^2}
\end{eqnarray*}
\begin{eqnarray*}\label{Aristotleenergy0}
E^{~\prime}=E-\vec{x}.R(\theta)\vec{f}+\frac{1}{2} k\vec{x}^2
\end{eqnarray*}
where $\vec{p}$ is a linear momentum, $h$ is an action, $\vec{f}$ is a force, $k$ is Hooke's constant, $E$ is an energy and
$j$ is an angular momentum.\\
 The coadjoint orbit is, in this case,
characterized by the two trivial invariants $h$ and $k$ (\ref{massaction0}), and by the
nontrivial invariants $s$ and $U$ given by:
\begin{eqnarray*}
s=j-\vec{p}\times\vec{q}+\frac{1}{2}m\omega\vec{q}~^2,~~~U=E-\frac{1}{2}k\vec{q}~^2
\end{eqnarray*}
where
\begin{eqnarray}\label{qfk}
 \vec{q}=-\frac{\vec{f}}{k}
\end{eqnarray}
 We see that the coadjoint orbit is $4-$dimensional. Let us denote it by
${\cal{O}}_{(h,k,s,U)}$.\\

The restriction $\Omega=(\Omega_{ab})$ of the Kirillov form to the orbit is then
\begin{eqnarray*}
\Omega=\left(
\begin{array}{cccc}
0&m\omega&k&0\\
-m\omega&0&0&k\\
-k&0&0&0\\
0&-k&0&0\\
\end{array}
\right)
\end{eqnarray*}
The modified symplectic form is explicitly given by
$$
\sigma=dp_i\wedge dq^i+\frac{1}{2}m\omega \epsilon^{ij}dq^i\wedge dq^j
$$
where $\vec{q}$ is given by relation (\ref{qfk}).\\
If $(y_a)=(p_1,p_2,q^1,q^2)$, the Poisson brackets are then explicitly given by
\begin{eqnarray*}
\{H,g\}=\frac{\partial H}{\partial p_i}\frac{\partial g}{\partial
q^i}-\frac{\partial H}{\partial q^i}\frac{\partial g}{\partial
p_i}+F_{ij}\frac{\partial H}{\partial p_i}\frac{\partial g}{\partial
p_j}
\end{eqnarray*}
where
\begin{eqnarray*}
F_{ij}=-m\omega{\epsilon_{ij}}
\end{eqnarray*}
This implies that
\begin{eqnarray*}
\{p_i,p_j\}=F_{ij}~,~\{p_i,q^j\}=\delta^{j}_i~,~\{q^i,q^j\}=0
\end{eqnarray*}
 Let the symplectic realization of the extended Aristotle Lie group on its coadjoint orbit be given by
$(\vec{p}^{~\prime},\vec{q}^{~\prime})=L_{(\theta,\vec{\eta},\vec{x},t})(\vec{p},\vec{q})$ .
By using relations (\ref{Aristotlemomentum0}), we have
\begin{eqnarray*}
\vec{p}^{~\prime}=R(\theta)\vec{p}-k[(R(\theta)\vec{q}+\vec{x})t-\vec{\eta}]+\frac{h}{r^2}\epsilon(\vec{x}),~\vec{q}^{~\prime}=R(\theta)\vec{q}+\vec{x}
\end{eqnarray*}
It follows that ($\vec{p}(t),\vec{q}(t))=D_{(0,0,0,0,0,t})(\vec{p},\vec{q})$ gives rise to
$$
\vec{p}(t)=\vec{p}-k\vec{q}~t,~~
\vec{q}(t)=\vec{q}
$$
The equations of motion are then
$$
\frac{d\vec{p}}{dt}=-k\vec{q},~~\frac{d\vec{q}}{dt}=0
$$
So with the central extension of the noncentrally extended of the two-dimensional Aristotle group, we have realized a phase space
where momenta do not commute and this noncommutativity is due to presence
of the magnetic field
\begin{eqnarray}\label{magneticfield}
F_{ij}=-m\omega{\epsilon_{ij}}=-eB\epsilon_{ij}.
\end{eqnarray}
Moreover, this phase space (the orbit) describes a spring submitted to a Hooke's force ($\vec{F}=-k\vec{q}$) which does not change the elongation
in time.\\~~\\
 \section{Conclusion}
 In this paper, we have proved that one can not construct noncommutative phase spaces by the coadjoint orbit method with the first and the second
central  extensions of the two-dimensional Aristotle group because symplectic structures obtained are canonical. But by considering the
 noncentrally extended Aristotle group and its corresponding central extension, we have realized partially noncommutative phase
spaces (only momenta do not commute).
In the first case, all the phase space coordinates do not depend on the time.  To overcome this situation, we have considered the central
extension of the above noncentrally extended Aristotle group. The phase space obtained in the latter case describes a spring submitted to a
Hooke's force ($\vec{F}=-k\vec{q}$) which does not change the elongation in time.  Furthermore, the noncommutativity of momenta is measured by a
term which is associated to the naturally introduced magnetic field (\ref{magneticfield}). Moreover, this case corresponds
to the minimal coupling of the momentum with the magnetic potential (\cite{ancilla}).
 
 \end{document}